\documentclass[runningheads]{llncs}
\usepackage{graphicx}
\usepackage{amsmath}
\usepackage{multirow}
\usepackage{booktabs}
\usepackage{threeparttable}
\usepackage[colorinlistoftodos]{todonotes}
\usepackage{subfigure}
\usepackage{makecell}
\usepackage{todonotes}
\usepackage{marvosym}

\begin{document}
\title{SkullEngine: A Multi-Stage CNN Framework for Collaborative CBCT Image Segmentation and Landmark Detection}

\titlerunning{SkullEngine}
% \author{Submission ID 100}

\author{Qin Liu\inst{1} \and Han Deng\inst{2} \and Chunfeng Lian\inst{1} Xiaoyang Chen\inst{1} Deqiang Xiao\inst{1} \and Lei Ma\inst{1} \and Xu Chen\inst{1} \and Tianshu Kuang\inst{2} \and Jaime Gateno\inst{2} Pew-Thian Yap\inst{1}\textsuperscript{(\Letter)} \and James J. Xia\inst{2}\textsuperscript{(\Letter)}}
\institute{
Department of Radiology and Biomedical Research Imaging Center (BRIC), University of North Carolina at Chapel Hill, NC, USA \\ 
 \and
Department of Oral and Maxillofacial Surgery, Houston Methodist Hospital, TX, USA\\
\email{qinliu19@email.unc.edu}, \email{ptyap@med.unc.edu}, \email{jxia@houstonmethodist.org}
}

\maketitle              % typeset the header of the contribution
\begin{abstract}
Accurate bone segmentation and landmark detection are two essential preparation tasks in computer-aided surgical planning for patients with craniomaxillofacial (CMF) deformities. 
Surgeons typically have to complete the two tasks manually, spending $\sim$12 hours for each set of CBCT or $\sim$5 hours for CT. To tackle these problems, we propose a multi-stage coarse-to-fine CNN-based framework, called SkullEngine, for high-resolution segmentation and large-scale landmark detection through a collaborative, integrated, and scalable JSD model and three segmentation and landmark detection refinement models. We evaluated our framework on a clinical dataset consisting of 170 CBCT/CT images for the task of segmenting 2 bones (midface and mandible) and detecting 175 clinically common landmarks on bones, teeth, and soft tissues. Experimental results show that SkullEngine significantly improves segmentation quality, especially in regions where the bone is thin. In addition, SkullEngine also efficiently and accurately detect all of the 175 landmarks. Both tasks were completed simultaneously within 3 minutes regardless of CBCT or CT with high segmentation quality. Currently, SkullEngine has been integrated into a clinical workflow to further evaluate its clinical efficiency. 

\end{abstract}

\section{Introduction}
Accurate bone segmentation and landmark detection are two fundamental tasks in preparing cone-beam computed tomography (CBCT) or computed tomography (CT)\footnote{For brevity, in the rest of the paper CBCT refers to both CBCT and CT. CBCT is more frequently used clinically.}
scans for use in computer-aided surgical simulation to treat patients with craniomaxillofacial (CMF) deformities. In current clinical practice, it takes at least a day and a half for a surgeon or a trained operator to manually perform both tasks to obtain the CBCT segmentation masks and landmark coordinates, which are time-consuming, labor-intensive, and error-prone. Therefore, there is an urgent need to develop a reliable automatic segmentation and landmark detection method for clinical use.

Automatic CMF bone segmentation and landmark detection are practically challenging due to the complex CMF anatomy, significant variations in appearance (especially in patients with severe deformities), and large image sizes (up to $768\times768\times576$). 
Most existing methods, including conventional methods \cite{gupta2015a,shahidi2014the,wang2016automated} and CNN-based methods \cite{lang2020automatic,minnema2019segmentation,torosdagli2018Deep,zheng20153d}, formulate the segmentation and the landmark detection as two independent tasks without considering their inherent relationship (e.g., landmarks usually lie on the boundaries of segmented bone regions). 

In recent years, CNN-based joint segmentation and landmark detection (JSD) approaches \cite{lian2020multi,zhang2020context} have been proposed to combine the two tasks via multi-task learning.
In \cite{lian2020multi}, the authors proposed a multi-task dynamic transformer network (DTNet) for concurrently segmenting mandible and detecting 64 landmarks.
In \cite{zhang2020context}, the authors proposed a context-guided multi-task fully convolutional network for jointly segmenting two bony structures (i.e., midface and mandible) and 15 boney landmarks.
However, these approaches have three major drawbacks that hinder them from being integrated into clinical practice: 1) they cannot detect large-scale landmarks (e.g., over 100 landmarks) due to the limited graphics processing unit (GPU) memory, 2) they cannot specifically refine segmentation in regions that are important for surgical planning (e.g., regions with thin bones), 3) they are non-scalable because the segmentation and landmark detection tasks are highly coupled in a single network. 
%Therefore, changing one task will easily affect the other's performance. Besides, in order to detect a new group of landmarks, we have to first modify the network structure and then retrain the whole model from scratch.
 
To tackle these issues, we propose a coarse-to-fine CNN-based framework, the SkullEngine, for high-resolution bone segmentation and large-scale landmark detection through a collaborative, integrated, and scalable JSD model and three refinement models. The goal of SkullEngine is to segment 2 bones (i.e., midface and mandible) and 175 landmarks (i.e., 66 for bones, 68 for teeth, and 41 for soft tissues) from a CBCT image. SkullEngine achieves this goal in two sequential stages: coarse and refinement. In the coarse stage, a scalable JSD model, a combination of 3D U-Net-based segmentation and landmark detection models, takes a down-sampled image as input for coarse segmentation and global landmark detection (i.e., all landmarks except the tooth landmarks). The tooth landmarks are not detected in this stage because they are close together in a small region and therefore need to detected in a higher resolution. In the refinement stage, based on the coarse segmentation mask and global landmarks achieved with the previous stage, region-of-interest volumes are cropped from the original CBCT image for further segmentation refinement and tooth landmark detection. 

A major technical contribution of the proposed SkullEngine is that our new JSD model is scalable and modular compared with previous JSDs. In addition, a major clinical contribution of SkullEngine is its clinical effectiveness. The use of SkullEngine can significantly reduce CBCT data preparation time from $\sim$12 hours, or CT from $\sim$5 hours, to 3 minutes for both clinically challenging tasks of high-resolution CBCT segmentation and large-scale landmark detection. We demonstrate in this study the accuracy of SkullEngine using a clinical dataset containing 92 CBCT scans and 78 CT scans of patients with CMF deformities. We have already integrated SkullEngine into a clinical workflow to continue to evaluate its efficiency in daily clinical practice.

\section{Methods}

\begin{figure}[t]
	\centering
	\includegraphics[width=11.8cm, height=5.3cm]{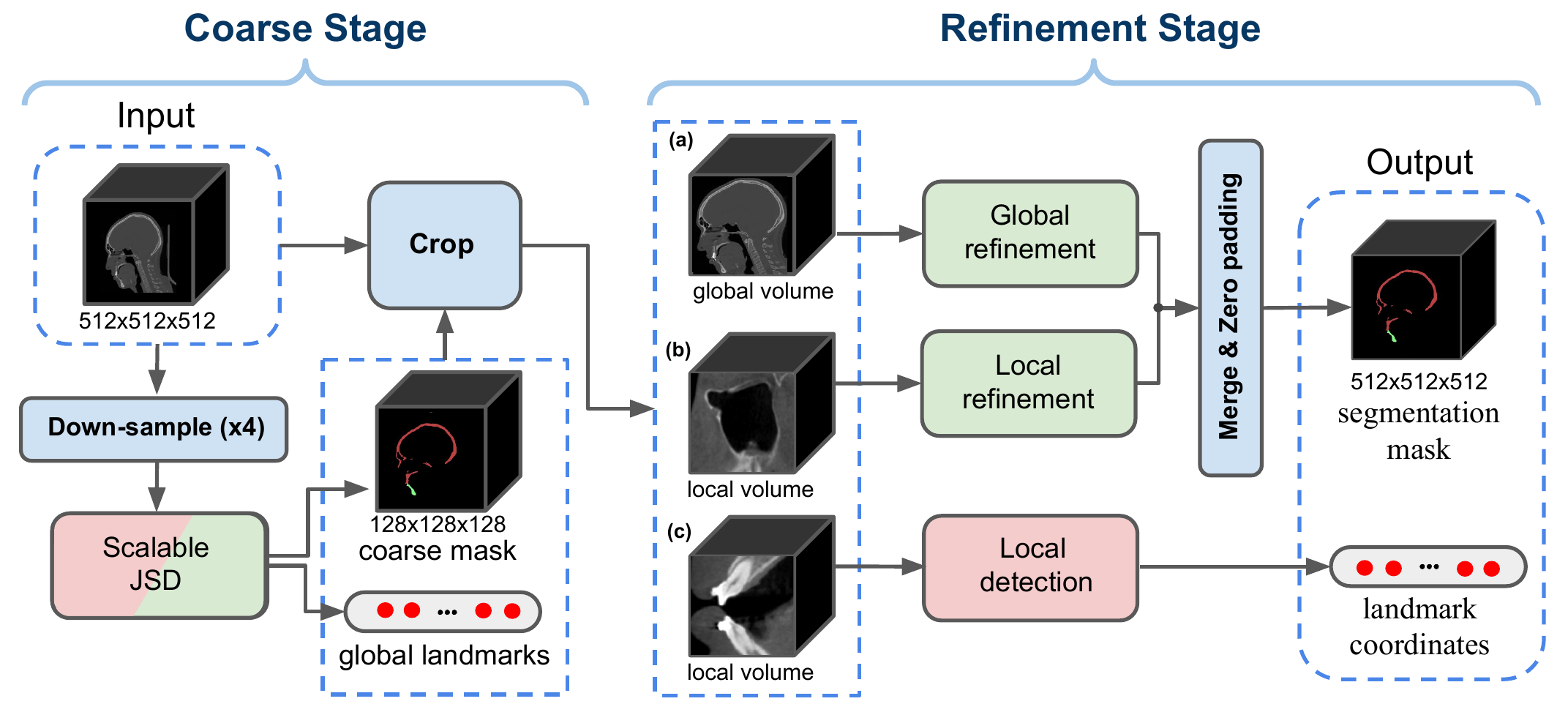}
	\caption{The framework of SkullEngine. The volumes (a) and (b) are cropped for segmentation refinement, while the volume (c) is cropped for tooth landmark detection. The size of each volume on the picture is just for illustration and may be various for different cases.}
	\label{fig:skullengine}	
\end{figure}
SkullEngine is a coarse-to-fine CNN-based framework that consists of two stages: first coarse then refinement (Fig. \ref{fig:skullengine}). In Section 2.1, we describe the coarse stage, in which a scalable JSD model is developed for coarse segmentation and global landmark detection (i.e., bony and facial landmarks). In Section 2.2, we describe the refinement stage, in which two segmentation models and one landmark detection model are developed for bone segmentation refinement and local tooth landmark detection, respectively. 

\subsection{Scalable JSD Model for Coarse Segmentation and Global Landmark Detection}
\begin{figure}[t]
	\centering
	\includegraphics[width=10cm, height=5.0cm]{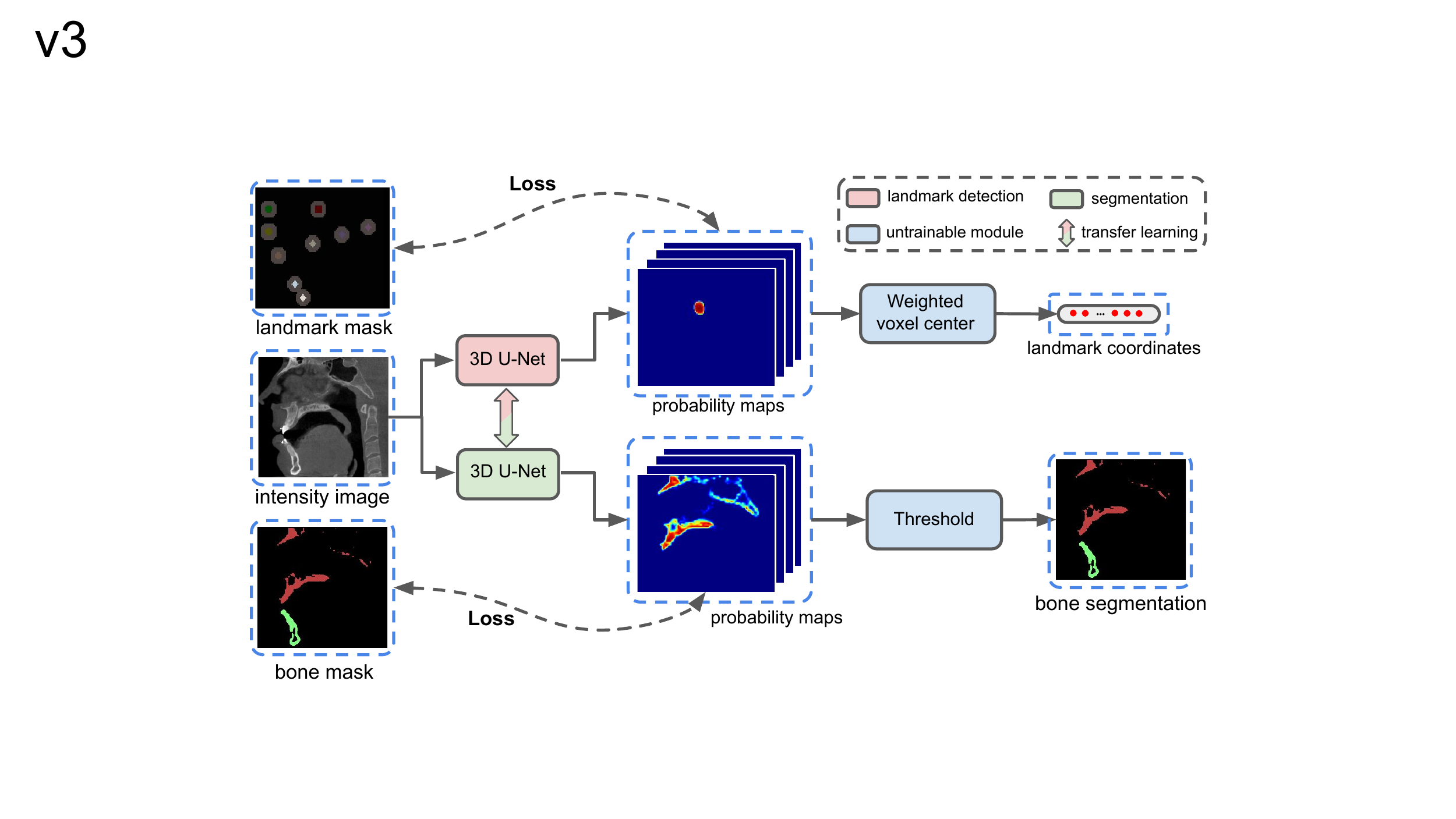}
	\caption{The training and inference framework for the scalable JSD model. The segmentation model and landmark detection model can be trained as a unified voxel-classification task. In the picture, we only show one segmentation model and one detection model. The transfer learning arrow between the two models means one model can be initialized by the weight from the other model instead of training from scratch.}
	\label{fig:seg_det}	
\end{figure}

Existing JSD models \cite{lian2020multi,zhang2020context} are non-scalable due to their ad-hoc network design for predefined segmentation and landmark detection tasks using different loss functions. 
With the breakthrough of U-Net \cite{cciccek20163dunet,ronneberger2015unet}, as well as its variants \cite{isensee2019nnunet,milletari2016vnet}, it is now possible to use U-Net as a building block to develop a scalable JSD model for unifying the segmentation and landmark detection tasks by taking both of them as a voxel-classification task. In addition, if the segmentation and landmark detection models have the same network structure, we can use transfer learning techniques \cite{Dodge2020fine} to train the models more efficiently. Inspired by these ideas, we propose a simple and scalable JSD model (Fig. \ref{fig:seg_det}). For segmentation, the ground truth is a voxel-level multi-class mask (Fig. \ref{fig:seg_det} bottom-left). For landmark detection, the ground truth is also a voxel-level landmark mask (Fig. \ref{fig:seg_det} top-left), which are generated from landmark-level coordinates.
We generate the landmark mask by assigning voxels to the label of a given landmark (e.g., the $i$-th label represents the $i$-th landmark) if these voxels belonging to the neighborhood of that landmark. We define the neighborhood of each landmark as a sphere with a predefined radius, which is a hyper-parameter we empirically set to 3 voxels.

\subsubsection{Construction of the Scalable JSD Model} Our JSD model consists of a segmentation model and two landmark detection models. The segmentation model is used for coarse bone segmentation of the midface and the mandible, whereas the two landmark detection models are used for detecting 66 bony and 41 facial global landmarks, respectively. However, the tooth landmark detection model is not included in the scalable JSD because the resolution of the down-sampled volume is too coarse to achieve any meaningful results.

\subsubsection{Training and Inference} In the training phase, the input images are down-sampled to a fixed resolution (e.g., 2.0\,mm$^3$ in our experiments) for training both the segmentation and landmark detection models. The segmentation model is first trained from scratch. The two landmark detection models are then initialized and fine-tuned using the parameters from the segmentation model. All are trained using the Focal loss function \cite{lin2017focal}. In the inference phase, the input image is down-sampled to the same resolution as in the training phase. We then run the three models independently for inference.

\subsection{Bone Segmentation Refinement and Local Landmark Detection}
The refinement stage aims to refine the bone segmentation results based on the coarse segmentation results and to detect the tooth landmarks that are undetectable in the coarse stage. As shown in Fig.~\ref{fig:skullengine} (right), the original image is cropped based on the coarse mask and global landmarks. Volume (a) in Fig.~\ref{fig:skullengine} is a global volume from cropped from the original image that contains the whole skull. The global refinement model uses a patch-based training and inference method for high-resolution segmentation in volume (a) (e.g., by cropping patches with a resolution of 0.4\,mm$^3$). Volume (b) in Fig.~\ref{fig:skullengine} shows a volume with thin facial bone. In fact, we need to crop two such volumes (for both the left and right facial bones) from the CBCT image using the paired right and left bony landmarks in the thin bone region as centers. Volume (c) in Fig.~\ref{fig:skullengine} is the tooth volume that is cropped also based on the bony landmarks. We train a tooth landmark detection model only in this region with a relative higher resolution (e.g., 0.8\,mm$^3$). During inference, we first crop the tooth volume that has the same size and resolution as the training patch based on the already detected bony landmarks, and then feed the cropped volume for tooth landmark detection. Finally, the segmentation mask obtained by the global and local refinement models are merged and zero-padded to the original size.

\section{Experiments and Results}
\subsection{Materials}

\begin{figure}[t]
	\centering
	\subfigure[Axial]{
	\includegraphics[width=2.7cm, height=2.7cm]{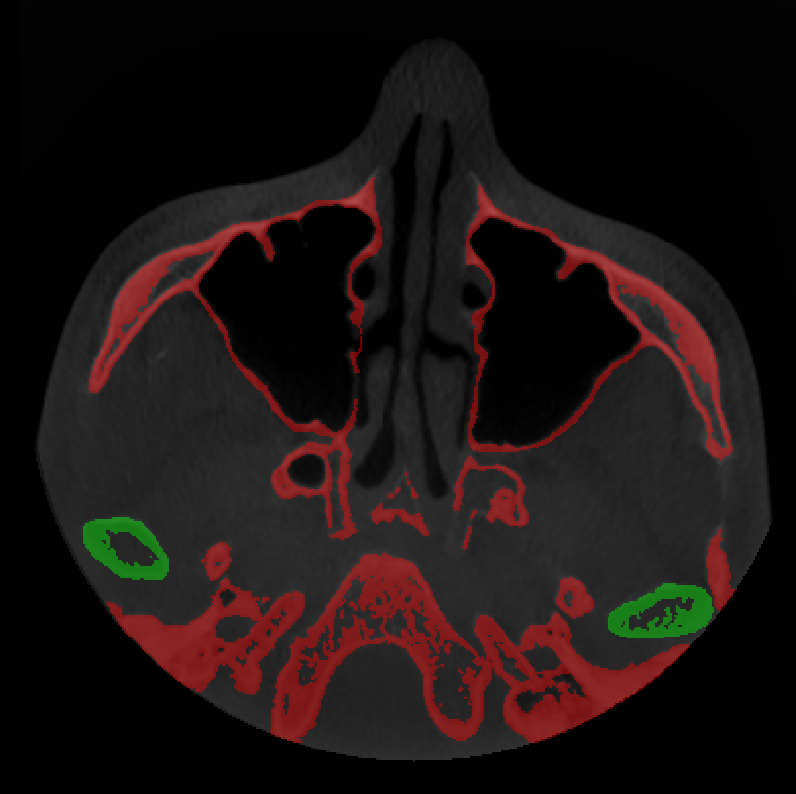}
	}
	\subfigure[Sagittal]{
	\includegraphics[width=2.7cm, height=2.7cm]{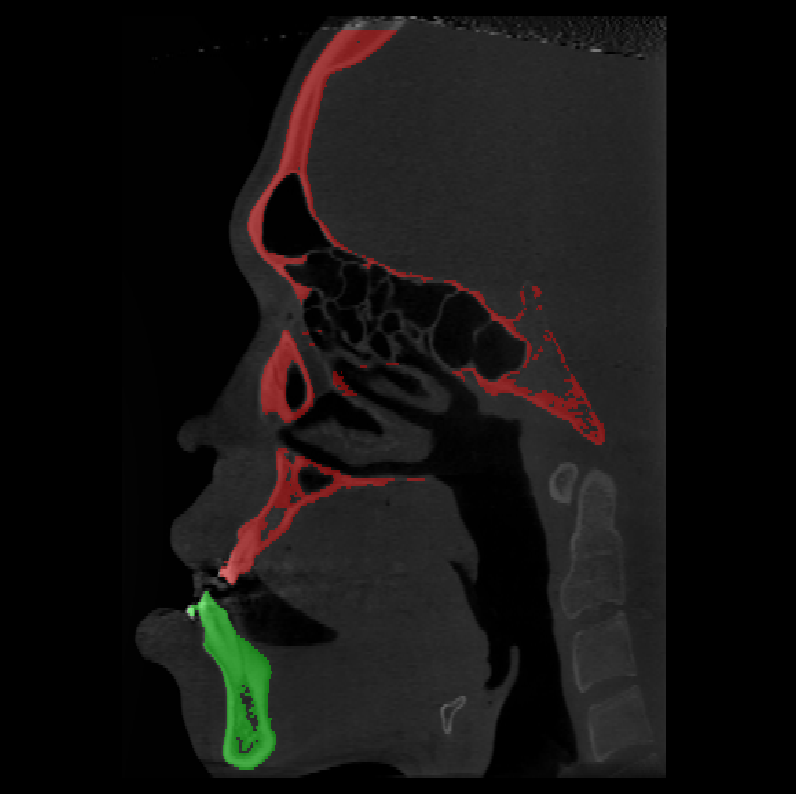}
	}
    \subfigure[Coronal]{
	\includegraphics[width=2.7cm, height=2.7cm]{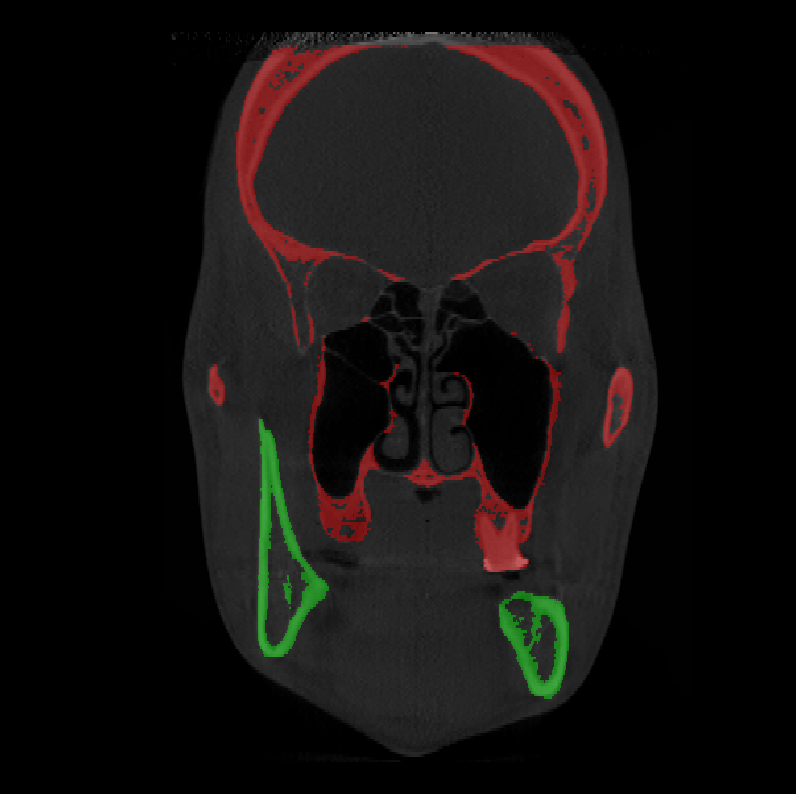}
	}
	\subfigure[Ground Truth]{
	\includegraphics[width=2.8cm, height=2.7cm]{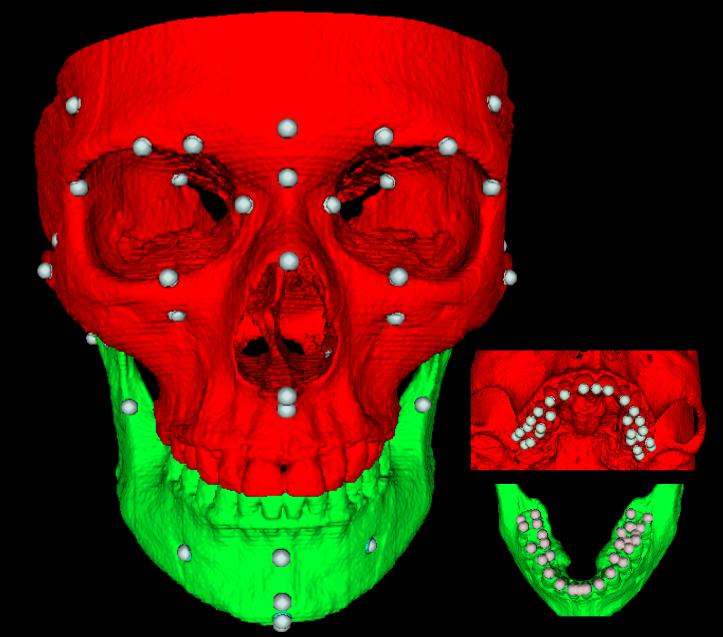}
	}
	\caption{An example of CBCT and the ground-truth landmarks. Red label represents the midface, and green represents the mandible. (a) Axial, (b) sagittal, and (c) coronal views, and (d) 3D view of the bony and tooth landmarks marked by spheres.}
	\label{fig:examplescans}	
\end{figure}

We evaluated the effectiveness of the proposed framework using 92 CBCT and 78 CT scans, which were randomly selected from our digital archive of 170 patients who had already undergone surgery in treating their CMF deformities (Table \ref{tab:dataset}). The study was approved by Institutional Review Board (Pro00013802). All personal information were de-identified prior to the study. Following clinical standard of care protocol, experienced CMF surgeons used currently available tools to generate segmentation masks and annotate landmark for all 170 scans as ground truth \cite{yuandesign2017}. The segmentation labels included two bony masks: midface and mandible. The 175 landmarks included 66 on both midface and mandible, 68 on both upper and lower teeth, and 41 on facial soft tissues. The average time of creating ground truth was 12 hours for each set of CBCT and 5 hours for each set of CT. Fig. \ref{fig:examplescans} shows a random example of CBCT scan and its ground truth.

\begin{table}[t]
	\caption{Summary of the CBCT/CT dataset.}	
	\newcommand{\tabincell}[2]{\begin{tabular}{@{}#1@{}}#2\end{tabular}}
	\begin{center}
		\begin{tabular}{c c c}
            \toprule
            & \textbf{CBCT/CT Dataset} \\
			\hline
			Number of scans & 170 (CBCT: 92, CT: 78) \\
			Dataset spliting & training 70\%, validation 10\%, testing 20\% \\
			Median spacing (mm$^{3}$) & 0.39$\times$0.39$\times$1.0 \\
			Median size & 512$\times$512$\times$418 \\
			Manual segmentations & midface, mandible (both including teeth) \\
			Number of landmarks per scan & 175 (bone: 66, teeth: 68, face: 41) \\
            \bottomrule
		\end{tabular}
	\end{center}
	\label{tab:dataset}	
\end{table}

We randomly divided the dataset into three groups: 119 scans (70\%) for training, 17 (10\%) for validation, and 34 (20\%) for testing. We applied a stratified sampling strategy to ensure each group included a balanced portion of CBCT and CT scans. The segmentation and landmark detection tasks were completed using our proposed coarse-to-fine SkullEngine framework. Finally, the archieved results were compared with two state-of-the-art competing methods: 3D U-Net \cite{cciccek20163dunet} and its upgraded variant based on PointRend \cite{kirillov2020pointrend}. We implemented the two methods for both segmentation and landmark detection tasks.

During the evaluation, the computational speed was calculated, starting from the input of CBCT image into SkullEngine until the output of the segmentation masks and landmark coordinates was completed. For segmentation task, we used Dice similarity coefficient (DSC), sensitivity (SEN), and positive prediction value (PPV) to evaluate the segmentation accuracy using means and standard deviations (SDs). For landmark detection task, we used root mean squared error (RMSE) and true positive rate (TPR) as metrics to evaluate landmark detection accuracy.

\subsection{Results}
\begin{figure}

	\centering
	\includegraphics[width=11.5cm, height=6.0cm]{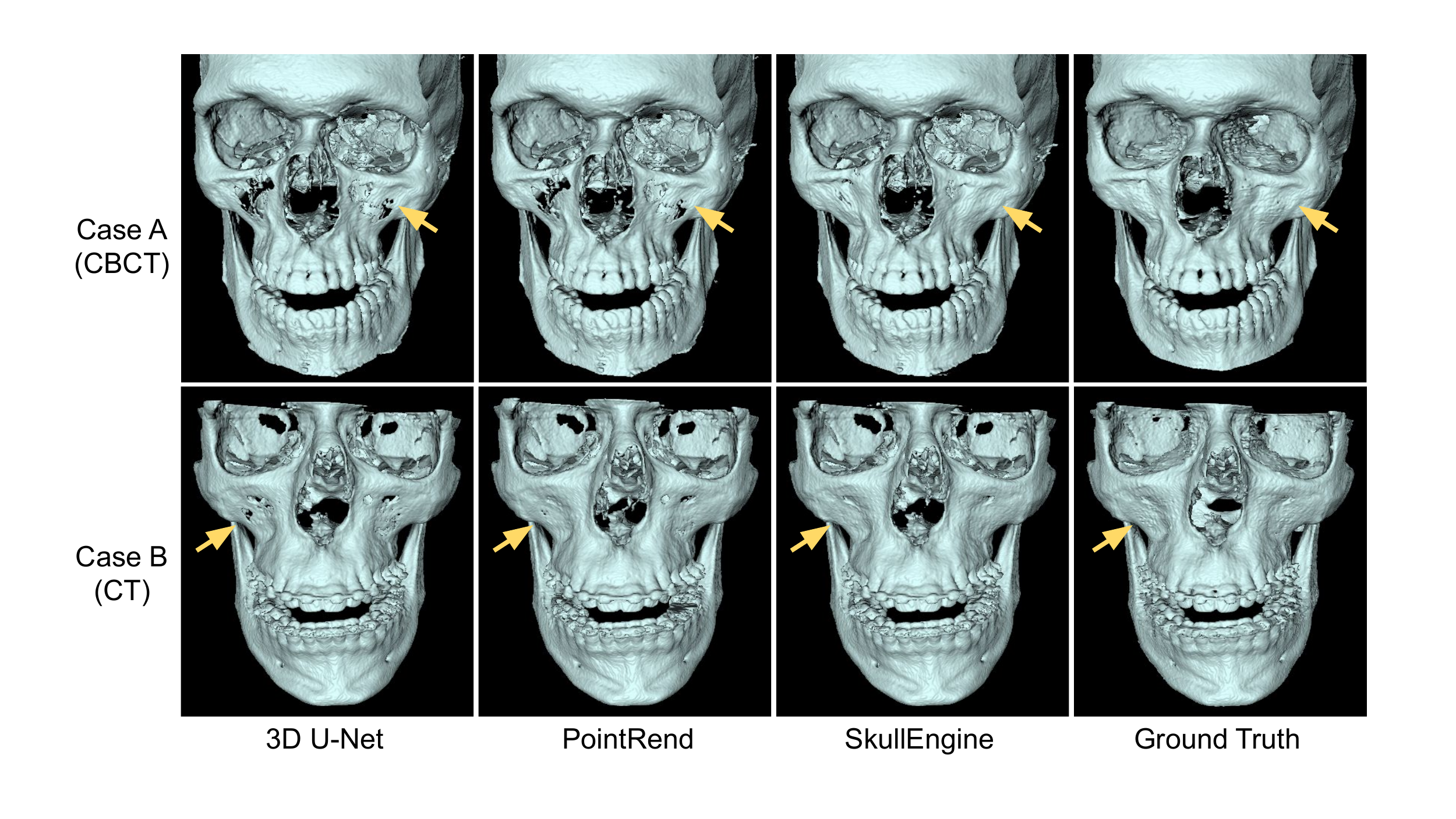}
	\caption{Comparison of segmentation results for two randomly selected cases. The thin bone areas are pointed by the yellow arrows.}
	\label{fig:seg_results}	
\end{figure}

For each testing dataset, both segmentation and landmark detection tasks were completed simultaneously within 3 minutes regardless of CBCT or CT. In contrast, a trained operator spent $\sim$12 hours for a set of CBCT and $\sim$5 hours for CT when the current clinical standard method was used. It indicates SkullEngine has a high degree of efficiency. 

\begin{figure}
	\centering
	\includegraphics[width=12.0cm, height=4.5cm]{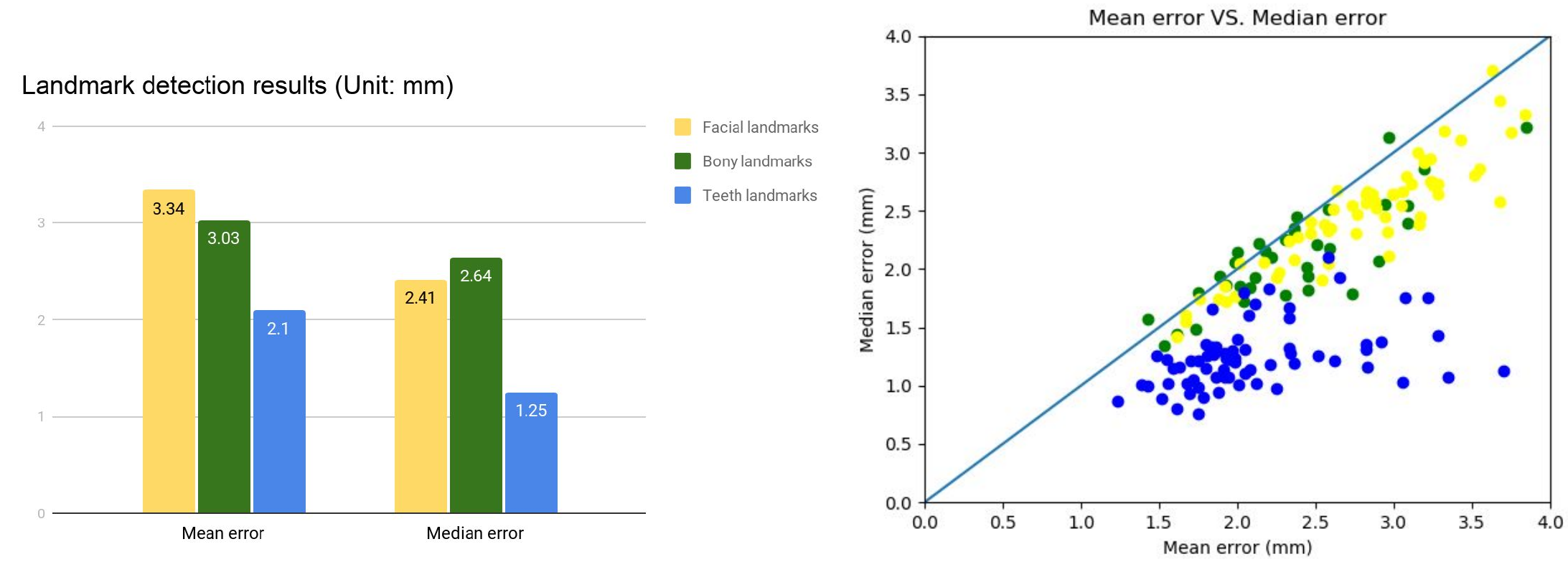}
	\caption{Results of landmark detection of all 175 landmarks using SkullEngine, including 66 bony, 68 teeth, and 41 facial landmarks.}
	\label{fig:det_results}	
\end{figure}

Table~\ref{tab:seg_results} shows the comparison results of segmentation accuracy among the three methods. The results clearly show that our SkullEngine is superior. Fig. \ref{fig:seg_results} shows 2 randomly selected examples. Note that in the thin bone areas are often misidentified as holes due to the low bony density. Since SkullEngine has a refinement model for segmentation, it is capable of solving the ``hole'' problem in the thin bone regions, unlike the other two competing methods.
Table~\ref{tab:det_results} shows the comparison results of landmark detection accuracy between the 3 methods. The results clearly show that our SkullEngine is more accurate. Fig.~5 (left) shows the detection results of SkullEngine on bones, teeth, and facial soft-tissue landmarks, respectively. Fig.~\ref{fig:det_results} (right) shows the error distribution of all 175 landmarks.

\begin{table}
	\caption{Comparison results of bone segmentation (mean$\pm$SD).}
	\begin{center}
	    \begin{threeparttable}
		\begin{tabular}{c | c | c | c | c | c | c | c}
		    \toprule
            \multirow{2}{*}{Method} & \multirow{2}{*}{\makecell[c]{Holes on \\ thin bone}} &
            \multicolumn{3}{|c|}{Midface} &
            \multicolumn{3}{|c}{Mandible} \\
            \cline{3-8}
            & & DSC (\%) & SEN (\%) & PPV (\%) & 
            DSC (\%) & SEN (\%) & PPV (\%) \\
            \hline
			3D U-Net \cite{cciccek20163dunet} & Yes & 87.9$\pm$7.4 & 83.2$\pm$9.5 & 89.2$\pm$7.9 & 
			89.4$\pm$4.1 & 91.7$\pm$7.1 & 93.7$\pm$4.9 \\
			PointRend \cite{kirillov2020pointrend} & Yes & 88.3$\pm$7.2 & 84.7$\pm$8.9 & 88.9$\pm$8.5 & 
			92.4$\pm$3.4 & 91.2$\pm$7.4 & 94.3$\pm$5.1 \\
			SkullEngine & No & \textbf{88.5$\pm$6.9} & \textbf{85.3$\pm$9.3} & \textbf{91.8$\pm$7.3} & \textbf{93.5$\pm$3.4} & \textbf{92.2$\pm$6.1} & \textbf{95.1$\pm$4.5} \\
            
			\bottomrule
		\end{tabular}
		\end{threeparttable}
	\end{center}
	\label{tab:seg_results}
\end{table}

\begin{table}
	\caption{Comparison results of landmark detection (mean$\pm$SD).}
	\begin{center}
	    \begin{threeparttable}
		\begin{tabular}{c | c | c | c | c | c | c }
		    \toprule
            \multirow{2}{*}{Method} & \multicolumn{2}{|c|}{Bony landmarks} & 
            \multicolumn{2}{|c|}{Tooth landmarks} & \multicolumn{2}{|c}{Facial landmarks} \\
            \cline{2-7}
            & RMSE & TPR (\%) & RMSE & 
            TPR (\%) & RMSE & TPR (\%) \\
            \hline
			3D U-Net \cite{cciccek20163dunet} & 3.17$\pm$1.79 & 96.7$\pm$3.1 & 2.58$\pm$3.03 & 
			97.3$\pm$4.1 & 3.46$\pm$3.31 & 96.4$\pm$4.2 \\
			PointRend \cite{kirillov2020pointrend} & 3.23$\pm$2.10 & 95.0$\pm$4.7 & 2.36$\pm$2.96 & 
			97.4$\pm$3.9 & 3.28$\pm$3.15 & 97.3$\pm$3.7 \\
			SkullEngine & \textbf{3.03$\pm$1.96} & \textbf{98.5$\pm$2.5} & \textbf{2.10$\pm$2.89} & \textbf{98.6$\pm$3.5} & \textbf{3.34$\pm$3.20} & \textbf{97.5$\pm$3.9} \\
			\bottomrule
		\end{tabular}
		\end{threeparttable}
	\end{center}
	\label{tab:det_results}
\end{table}

\subsection{Implementation details}
We conducted our experiments on a standard workstation equipped with Intel dual-Xeon E5 CPUs, and a single NVidia Titan XP GPU with 12 GB memory. The SkullEngine was implemented and trained using Python 3.7 and Pytorch 1.7.

\section{Conclusion}
In this work, we have proposed a clinical practical framework for high-quality segmentation and large-scale landmark detection from skull CBCT/CT scans. Our multi-stage framework, the SkullEngine, collaboratively integrates segmentation and landmark detection models to maximize the overall performance. The experimental results showed its superior accuracy when compared to the state-of-the-art methods. The results also showed a significant reduction in labor and time spent on CBCT data preparation from ~12 hours to less than 3 minutes with high-quality segmentation results, highlighting the practical value of SkullEngine. Currently, we have integrated SkullEngine into our clinical workflow for further evaluating its clinical efficiency.

\section*{Acknowledgement}
This work was supported in part by United States National Institutes of Health (NIH) grants R01 DE022676, R01 DE027251, and R01 DE021863.

\bibliographystyle{splncs04}
\bibliography{main}

\end{document}